# Deterministic Intra-Vehicle Communications: Timing and Synchronization

Hamid Gharavi and Bin Hu
Advanced Network Technologies
National Institute of Standards and Technology
Gaithersburg, USA
Emails: [Gharavi, bhu]@nist.gov

*Abstract*— As we power through to the future, in-vehicle communications' reliance on speed is becoming a challenging predicament. This is mainly due to the ever-increasing number of electronic control units (ECUs), which will continue to drain network capacity, hence further increasing bandwidth demand. For a wired network, a tradeoff between bandwidth requirement, reliability, and cost-effectiveness has been our main motivation in developing a high-speed network architecture that is based on the integration of two time-triggered protocols namely; Time Triggered Ethernet (TT-E) and Time Triggered Controller Area Network (TT-CAN). Therefore, as a visible example of an Internet of Vehicles technology, we present a time triggered communication-based network architecture. The new architecture can provide scalable integration of advanced functionalities, while maintaining safety and high reliability. To comply with the bandwidth requirement, we consider high-speed TT-Ethernet as the main bus (i.e., backbone network) where sub-networks can use more cost-effective and lower bandwidth TT-CAN to communicate with other entities in the network via a gateway. The main challenge in the proposed network architecture has been to resolve interoperability between two entirely different time-triggered protocols, especially in terms of timing and synchronization. In this paper, we first explore the main key drivers of the proposed architecture, which are bandwidth, reliability, and timeliness. We then demonstrate the effectiveness of our gateway design in providing full interoperability between the two time-triggered protocols.

*Keywords—Internet of Vehicles (IoV), Time Triggered Ethernet (TT-E), Time Triggered Controller Area Network (TT-CAN), Electronic Control Units (ECUs), Intra Vehicle Communications*

## I. INTRODUCTION

A rapid growth in electric vehicles and Advanced Driver Assistant Systems (ADAS) is transforming future cars into large mobile data centers. The most important example of such a transformation is the self-driving car technology, which requires a large number of smart devices and sensors that can guarantee much better safety than manual driving. This, together with the increasing scale and scope of data generated from a vehicle with multiple ECUs, represents a tremendous challenge for future intelligent transportation under the auspices of the Internet of Vehicle (IoV).

Currently, vehicle electronics consist of several "sub-systems" or "domains" where each has its own control units, such as mechanical, electrical, or computer controls. The data generated by these sub-systems can vary with respect to bandwidth, reliability, and latency requirements. There are several advanced intra-vehicle communications standards, such as TT-CAN [1] and FlexRay [2, 3]. In particular, TT-CAN, which is an extended version of CAN [4], has been one of the most cost-effective communications deployed in today's vehicles [4]. While TT-CAN can fulfill the short term need for time-critical in-vehicle communications, its limited bandwidth is insufficient to respond to the ever increasing demand for more bandwidth. Such a demand has been further intensified in recent years by the introduction of more bandwidth hungry services, such as infotainment, camera-based Advanced Driver Assistance Systems (ADAS), advanced Laser detection and ranging (LADAR) scanners for crash detection and prevention [5], as well as future advanced 3-dimensional range video [6].

As a long-term solution, Ethernet technology, which is capable of providing a much higher bandwidth, embraces a logical transition towards IoV [7-9]. Currently, there are a few Ethernet-based protocol standards for in-vehicle communications [10]. While these protocols have undergone massive advancements, the reliability of the network to respond to emergency situations in a timely manner has been the main requirement.

Therefore, to keep up with the increasing integration of multiple sub-networks and cope with the staggering amount of video streaming for both safety and entertainment, in-vehicle communication would require not only faster, but highly reliable Ethernet-based networks, such as Time-triggered Ethernet (TT-E) [11]. On the other hand, transitioning from non-Ethernet-based TT-CAN to TT-E is a huge undertaking in terms of cost as most existing in-vehicle components are generally developed for CAN. Therefore, as a comprise between cost and capacity, in this paper we present a new network architecture, which is based on two time-triggered protocols. The main objective of integrating the two protocols is to expand the bandwidth and yet reduce the overall cost as much as possible. The main challenge, however, has been to achieve interoperability between the two time-triggered protocols. The proposed architecture provides a scalable integration of both protocols with advanced functionalities, while ensuring safety and reliability.

After a brief overview of the existing protocols, we provide greater details of the time protocols that have been used in our architecture. Section II presents a brief review of some of the most popular in-vehicle protocols. We discuss the potential of Ethernet based systems, especially the TT-E standard, for advanced in-vehicle communications. In Section III, we present our proposed network architecture, which includes our novel gateway design for achieving interworking and synchronization between the two time-triggered protocols, namely TT-CAN and TT-E. Finally, Section IV presents the simulation results of the integrated time protocols followed by the conclusion.

## II. TIME TRIGGERED INTRA-VEHICLE COMMUNICATIONS

Currently there are a number of intra-vehicle communication protocols such as Controller Area Network (CAN) [4], LIN (Local Interconnect Network) [12], FlexRay [2, 3], Media Oriented Systems Transport (MOST) [13] that have been used in today's vehicle. CAN however, is the most widely utilized technology. It is available in different forms, such as low and high-speed CAN, with Flexible Data Rate. Low-speed CAN has a data rate of 40 Kbps to 125 Kbps while the high-speed CAN offers bandwidth from 40 Kbps to 1 Mbps (depending on the length of the cable). In addition, the high-speed Time-triggered CAN (TT-CAN) offers bandwidth from 40 Kbps to 1 Mbps (depending on the length of the cable) [1]. The TT-CAN has been extensively deployed in today's automobile. Bear in mind that time triggered protocols aim to provide reliable distributed computing and networking for in-vehicle communication systems. In a time-triggered system, the activities are initiated periodically to ensure a high level of determinism. However, the increasing number of ECUs, smart sensors, and other data-generating devices in modern vehicles, would require a much greater bandwidth that the TT-CAN cannot provide.

To meet the demand for greater bandwidth, FlexRay was introduced, which can support data rates of up to 10 Mbps. Some of the key characteristics of the FlexRay protocol are; synchronous and asynchronous frame transfer, guaranteed frame jitter and latency during synchronous transfer, single as well as multi-master clock synchronization, prioritization of frames during asynchronous transfer, error detection and signaling, time synchronization across multiple networks, and scalable fault tolerance [2].

TT-Ethernet (TT-E) is another standard, which is designed to expand traditional Ethernet with services to fulfill the requirements of deterministic, time-critical, and safety-related applications. TT-E is based on the AS6802 standard [11], which has been developed for integrated systems and safety-related applications primarily in aerospace, industrial controls, and automotive applications. It is a Layer 2 Quality-of-Service (QoS) enhancement, which is designed to expand traditional Ethernet-based networks with services to fulfill the requirements of deterministic and time-critical traffic, as well as other types of non-critical traffic. This is achieved by providing different traffic classes in parallel, namely Time-Triggered (TT), Rate-Constrained (RC), and Best Effort (BE). For instance, a time-triggered message, which has the highest priority transmission, is scheduled over the network at pre-configured time instances. The generation triggered time, delay, and precision of time-triggered messages are pre-set and guaranteed. The rate-constrained messages have less strict determinism and real-time requirements. Rate-constrained messages have a predefined bandwidth with temporal deviations of well-defined bounds. The best-effort messages are similar to traditional ethernet traffic where no delay is guaranteed. These messages use the residual bandwidth of the network and have the lowest priority. TT-E, due to its important features such as low latency, reliability, and higher bandwidth, has been selected as the backbone network in our proposed architecture.

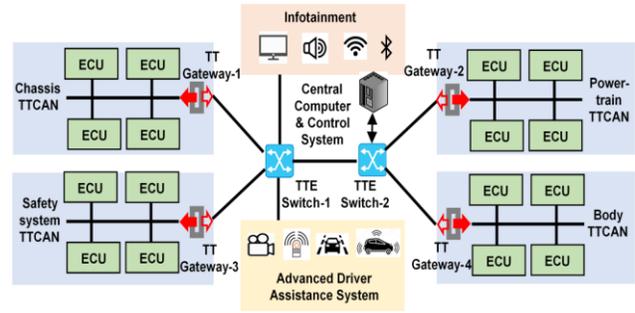

Fig. 1. A simplified in-vehicle time triggered communication system.

## III. PROPOSED NETWORK ARCHITECTURE

Today's vehicles are becoming more complex than ever mainly due to more advanced requirements for safety regulations, luxury conveniences, computer-based diagnostics, and a vast array of power accessories supported by each car. These requirements have raised a need for more Electronic Control Units (ECUs) in cars. These ECUs not only make monitoring easier, but also report back to the driver if something is wrong. An ECU is a computer module that communicates with the sensors deployed in a car and controls various electrical functions. There are several ECUs in modern cars that control different operations ranging from monitoring engine performance to controlling electronic accessories. Some common ECUs include powertrain control module, chassis control module, advanced driver assistance system, safety system, infotainment module, and comfort module. These ECUs exchange information with each other using various in-vehicle networking technologies. Amongst the most critical in-vehicle components are the advanced driver assistance system and its extension to all aspects of the drive towards "self-driving" cars. The advanced driver-assistance systems (ADAS) mainly focus on enhancing the driving experience by offering technologies to avoid accidents and collisions. These technologies help in detecting potential problems and alerting the driver. The ADAS relies on various types of cameras (e.g., right, left, front, and rear cameras), LADAR, and lane departure warning (LDW)/traffic signal recognition (TSR) systems. The cameras allow the driver to easily detect approaching pedestrians, vehicles, or cyclists with the help of a screen. For instance, the LDW/ TSR mechanisms are used to alert the driver when a vehicle begins to move out of its lane without any prior signal indication. These mechanisms aim at minimizing accidents by relying heavily on the ADAS to guarantee safety of vehicles and passengers and to perform corrective actions such as returning the vehicle to its lane or emergency braking, especially in the case of Self-Driving Automation.

With the growing number of ECUs and further expansion of ADAS for self-driving, the major challenge is how to provide enough bandwidth, as well as ensuring reliability of the in-vehicle communication network. TT-E not only can provide a huge bandwidth, but also offers deterministic communications for safety related sensors and ECUs. The main drawback of the TT-E is the deployment cost (e.g., switches and end-system) compared with the bus-based TT-CAN technology.

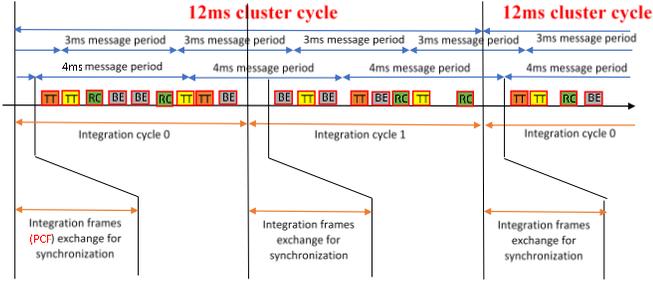

Fig. 2. An example of message periods, cluster cycle and integration cycles.

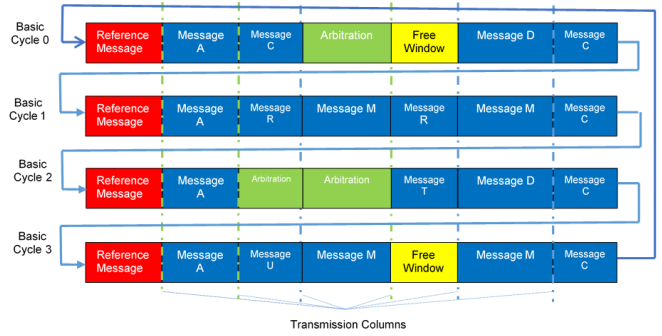

Fig. 3. TTCAN system matrix.

Furthermore, many traditional ECUs have already been designed for deployments in the TT-CAN environments. Therefore, as a comprise solution between cost and capacity, in this paper we a present a network architecture, which uses TT-E as the backbone network with multiple TT-CANs as sub-networks. Fig. 1 shows a simple example of the proposed architecture showing interworking between the two time-triggered protocols: TT-E and TT-CAN, via the design of a suitable gateway that can achieve interoperability between two entirely different time-triggered protocols.

*A. TT-Gateway Design*

The main functionality of the gateway is to allow interworking between two distinctly different time triggered protocols. This would require developing strategies to solve incompatibilities in terms of physical infrastructure, timing, and synchronization. To explore this further, we first provide more detailed information about the timing protocols of TT-E and TT-CAN.

For TT-E synchronization, the AS6802 protocol uses dedicated messages called protocol control frames (PCF) to establish and maintain system-wide clock synchronization amongst all the nodes at regular intervals (e.g., 1 millisecond), which is called integration cycles. The Synchronization entities consist of Compression Master (CM) and Synchronization Master (SM) or synchronization clients (SC), which are selected based on the system architecture. Generally, switches are configured as CM and end-systems as SM. Switches and end-systems that are not configured as synchronization master or compression master will be configured as the synchronization client.

The synchronization is achieved in two steps during normal operation mode (as shown in Fig. 11 of [11]). In the first step the synchronization masters send a PCF message (i.e., a short Ethernet frame: 64 bytes) to the compression masters (CM). The CMs then compute the average value using the relative received times of these PCFs. In the second step, the CMs then send out a new PCF to the SMs and SCs. More specifically, a TT-E switch, as the Compression Master (CM), generates a global time based on the PCF frames received from the Synchronization Masters (SMs) and distributes it to all Synchronization Masters and Synchronization Clients (SCs). In the two-step synchronization, the PCF frame (i.e., exchanges between SM and CM) is known as the integration frame, whereas the synchronization procedure for dispatching these integration frames (sending and compressing) within a configurable period is defined as the integration cycle. Fig. 2 shows the relationship between the integration cycle and the cluster cycle, where the overall cluster cycle in TT-Ethernet consists of multiple integration cycles, ranging from 0 to max_integration_cycle-1 [11]. On the other hand, the cluster cycle comprises the least common multiple of message periods, as depicted in Fig. 2.

Unlike TT-E, TT-CAN is a serial bus network and its messages are broadcast on the bus (instead of using sender and receiver's addresses) where they can be picked up by interested receivers according to the message's unique identifier. As a TDMA based scheme, TT-CAN divides the timeline between two consecutive reference messages, called the basic cycle, into time slots (windows). As shown in Fig. 3, a system matrix can be constructed by grouping a number of basic cycles. A basic cycle comprises several time windows of different sizes where a time master can send a reference message to every node to achieve synchronization. A time window may be an exclusive window, an arbitration window, or a free window. Exclusive windows are mainly considered for periodic messages without having to compete for network access (i.e., deterministic). An arbitrating window is for event triggered messages and a free time window is reserved for future extensions. A TT-CAN node, however, does not need to know the whole system matrix (only the information in each message) and based on the reference message, nodes can update their local time slots for transmission and reception of their data messages.

There are two possible timing levels in TT-CAN. In the first level (level-1) a time-triggered operation is carried out based on the reference message from a time master while an independent clock runs in each node. On the other hand, level-2, with the support of global time and a continuous drift correction in each node, provides higher synchronization quality. In our model, level 2 timing is considered where the TT-CAN section of the gateway operates as the time master in the TT-CAN network. Under level-2 timing, each node uses a cyclically incrementing counter as its Local-Time, which is decided by the Network Time Unit (NTU). The local time contains at least 19 bits where the three least significant bits represent a fraction of the NTU and is incremented in the unit of $NTU/2^n$. The length of the NTU is configurable and generated locally based on the local Time Unit Ratio ($TUR$) where $NTU = TUR \cdot t_{sys}$ and $t_{sys}$ is the local clock period. It should be noted that TUR is a non-integer value and can be adjusted accordingly for continuous drift correction [1]. More specifically, TUR can be adjusted based on the difference between local time and the time transmitted by the gateway (i.e., time master).

As shown in Fig. 3, at the beginning of each basic cycle a time master, which is the gateway in our proposed architecture, sends a reference message to every node within its sub-network and start its Cycle_Time (CT). Based on this message, nodes can update their local time slots for the transmission and reception of their messages. When the CT reaches a predefined value: $T_{cycle}$ (the length of the basic cycle), the time master resets its CT, starts a new basic cycle by sending a new reference message to the TT-CAN sub-network. On the other side, when receiving a reference message every node starts its new basic cycle and resets its CT. The value of the local time is saved as the Sync_Mark at the sample point of the Start-Of-Frame (SOF) bit of each message. The Sync_Mark of a reference message is defined as the Ref_Mark.

Fig. 4 shows the level-2 timing process for obtaining the cycle time and global time. As shown, the difference between a node's local time and Ref_Mark is the Cycle_Time, which is reset at the beginning of each basic cycle as soon as Ref_Mark is captured [14]. It should be noted that Global_Time is employed only in level-2 as the reference for the synchronization and calibration of all the local times to the time master's clock, hence providing more fine-grained synchronization. This is generated by the time master and transmitted in the reference message as Master_Ref_Mark to all nodes. The TT-CAN nodes derive their Global_Time by summing their Local_Time and their Local Offset (see Fig. 4). Local_Offset is the difference between the Ref_Mark in Local_Time and the Master_Ref_Mark in Global_Time, caused by the difference between local nodes' NTU and the time master's NTU. In the time master, the Local_Offset is zero. By comparing the differences between two consecutive Master_Ref_Marks (measured in time master's global NTUs and received in reference messages) and two consecutive Ref_Marks (measured in local NTUs), local nodes can derive the clock speed difference and compensate the drift by updating their TURs as follows: $TUR = df \cdot TUR_{previous}$, where drift factor $df = \frac{Ref\_Marks - Ref\_Marks_{previous}}{Master\_Ref\_Marks - Master\_Ref\_Marks_{previous}}$. Local nodes will then update local NTUs after this drift compensation (see Fig. 4) and calibrate local time base to the time master's time base.

### B. Integrated TT-E and TT-CAN

Figure 5 shows an example of the interworking between TT-E and TT-CAN protocols via a TT-gateway. The function of the gateway is to perform timing and synchronization between the two protocols. Specifically, the TT-E synchronization master in the TT-gateway synchronizes with the TT-E switch (compression master) periodically. Meanwhile the TT-CAN time master in the TT-gateway updates its clock to the local TT-E clock at the beginning of each basic cycle to achieve synchronization with the compression master (as seen in Fig. 5). Under these conditions, a TT-E switch, as the Compression Master (CM), generates a global time that is based on the PCF frames received from SMs, which is then distributed to all SMs and SCs. The synchronization process within the integration cycle operates inside a configurable period where TT-E clocks in TT-gateways can be updated. On the other hand, before sending the Reference (REF) message to the local TT-CAN network, the TT-CAN time master in the TT-gateway updates its clock to the local TT-E clock at the beginning of every basic cycle. The period in which the TT-CAN can be updated depends on the total amount of aggregated TT-CAN data that can be generated by multiple ECUs in the TT-CAN subnetwork. Therefore, due to the much larger bandwidth of the TT-E backbone network, such a time interval should be much larger than the TT-E integration cycle. As an example, in our simulation, the TT-CAN is updated every 0.024576 seconds (configurable) with respect to the TT-E integration cycle of 0.003.

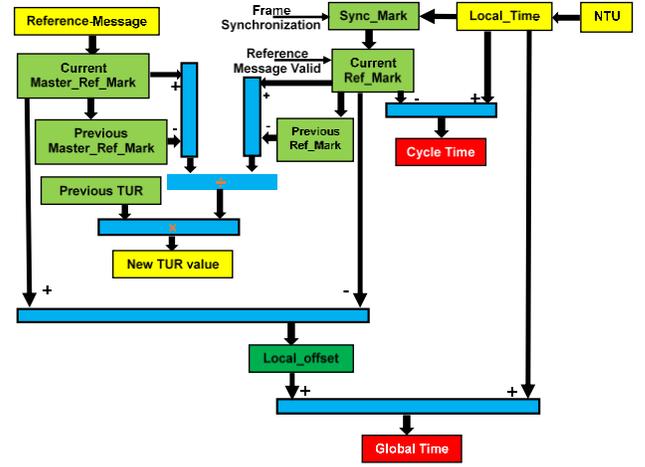

Fig. 4. Level-2 timing process: Cycle_Time, Global_Time, Local_offset and New TUR.

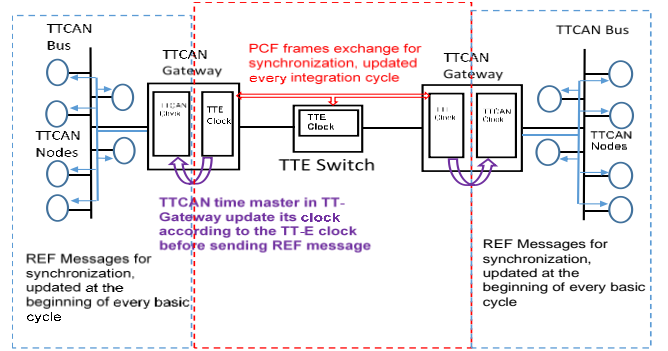

Fig. 5. An example of interworking between TT-E and TT-CAN protocols via a TT-Gateway.

Without an external clock, the TT-CAN time master can keep its TUR and NTU constant. However, in our system, the time master (the TT-CAN clock in the TTCAN gateway) will update itself with the TT-E clock in order to achieve full synchronization in the proposed integrated TT-E and TT-CAN network. At the end of the basic cycle, the TT-CAN clock in the TT-gateway first updates its Local_Time by incorporating the difference between itself and the TT-E clock; $T_{gap}$, where

$$\text{Local\_Time} = \text{Local\_Time}_{previous} + T_{gap}.$$

The new Local_Time is saved as the Master_Ref_Mark. It then updates its

$$df = \frac{Local\_Time_{previous} - Master\_Ref\_Marks_{previous}}{Master\_Ref\_Marks - Master\_Ref\_Marks_{previous}},$$
$$TUR = df \cdot TUR_{previous}$$

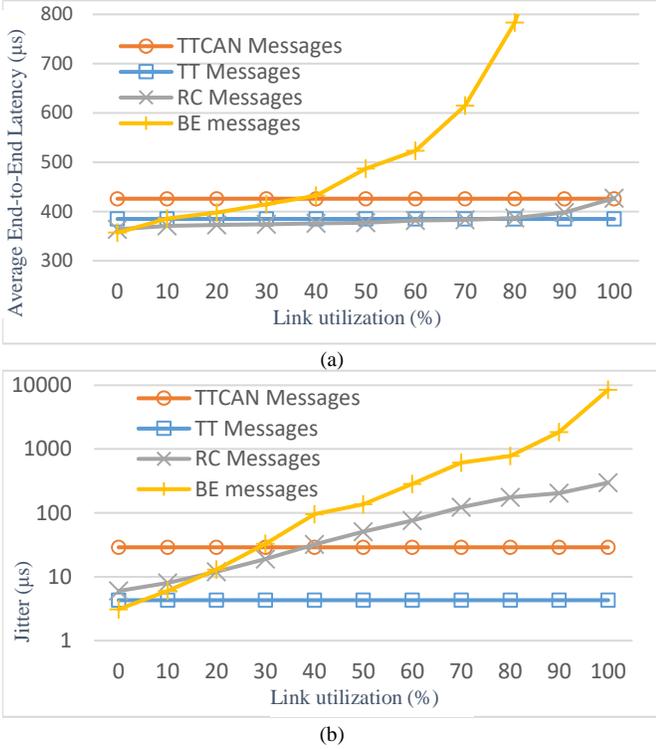

Fig. 6. The influence of link utilization on (a): average end-to-end latency performance and (b): the jitter performance of all traffic.

and NTU, calibrating their time base to that of the TT-E backbone network. Furthermore, it sends a new reference message to TT-CAN sub-network and resets its Cycle_Time to $T_{gap}$ (not zero as before). This means that the TT-CAN clock in TT-gateway will be re-synchronized with the TT-E clock after $T_{cycle} - T_{gap}$ seconds.

When receiving reference messages, the local node in the TT-CAN sub-network will first update its $df$ as

$$df = \frac{Local\_Time - Ref\_Marks_{previous}}{Master\_Ref\_Marks - T_{gap} - Master\_Ref\_Marks_{previous}}.$$

It then updates $TUR$ and NTU. Note that the time master has updated its Local_Time and Master_Ref_Marks by incorporating the $T_{gap}$ between the TT-CAN sub-network and the TT-E backbone network. The local node then updates its Local time to Master_Ref_Marks + τ (τ is the transmission delay between the local node and the time master) and saves it as Ref_Mark, to achieve synchronization with the time master.

## IV. SIMULATION

In this section, we evaluate integrated time triggered communication in the system topology of a simplified in-vehicle communication network shown as in Fig. 1, where video, audio, control, and inter-gateway traffic are categorized as BE traffic, RC traffic and TT traffic for investigation.

As shown in Fig. 1, the integrated in-vehicle communication network consists of two TT-E switches and 4 TT-Gateways producing a TT-E backbone network. Each TT-Gateway interconnects a TT-CAN sub-network with the TT-E backbone network. High-speed devices, such as ADAS and infotainment modules, are directly connected to one of the TT-E switches, sending TT traffic, RC traffic, and BE traffic to the control system through TT-E links. All TT-E links are bidirectional 100 Mbps links. On the other hand, low-speed devices, such as ECUs are connected to TT-CAN sub-networks with a 1 Mbps bandwidth. In our simulation, TT-CAN messages between ECUs are first sent to the local TT-Gateway and converted to TT messages, which will be sent to destination TT-Gateways via TT-E switches. A destination TT-Gateway first converts them back to TT-CAN messages before forwarding them to the destination ECUs or the central control unit.

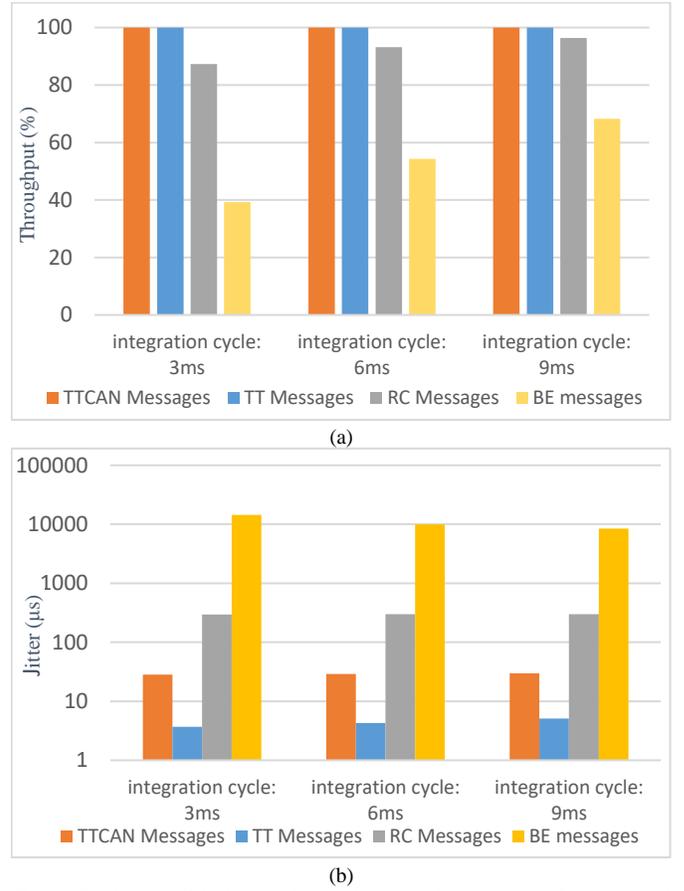

Fig. 7. The impact of the integration cycle on (a): throughput performance and (b): jitter performance.

For simplicity, the payload of the TTCAN traffic is set at 8 bytes, the payload of the TT traffic is 50 bytes, the payload of the RC traffic is 100 bytes, while the payload of the BE traffic is 500 bytes. The drifts of TT-E clocks and TT-CAN clocks are configured as 200 parts per million (ppm). The propagation delay of the TT-E backbone network is 100 ns per link, while the propagation delay of the TTCAN bus is set to 100 ns (5 ns per meter propagation delay and a maximum cable length of 40 meters on the CAN bus with 1 Mbps bandwidth [4]).

In a TT-E network, virtual links are used as logical connections to route traffic from a sender to one or more receivers and are pre-configured during the scheduling process. TT-E switches use pre-defined forwarding tables to concatenate virtual links and create tree structures with one sender as root and multiple receivers as leaf nodes [15]. Fig. 6 depict the average end-to-end latency and jitter performance of the in-vehicle time triggered communication network under the influence of different link utilizations caused by varying amounts of traffic. The TTCAN traffic, as a special and small part of TT Traffic in the TT-E backbone network, is measured with TT traffic but listed separately. Due to synchronization protocols and the link reservation mechanism in both TT-E backbone networks and TT-CAN sub-networks, TTCAN traffic and TT traffic are able to achieve a constant average end-to-end latency with very low jitter, despite the increasing link utilization. Their low jitters are mainly due to the clocks' drift.

Since TTCAN traffic experiences extra links with a lower timing resolution in TT-CAN sub-networks, it has a higher jitter (28.97 μs) and latency (426 μs) compared with TT traffic (i.e., jitter of 4.3 μs and latency of 385 μs). The jitter caused by the TT-CAN clocks' drift is higher than the jitter caused by the TT-E clocks' drift. The basic cycle (e.g. 0.024576 s) in TT-CAN is much longer than the integration cycle (e.g. 0.024576 s) in TT-E, resulting in more drift in TT-CAN clocks than in TT-E clocks before clock correction. The latency and jitter performance of the RC traffic degrades because of increasing link utilization. However, it is limited to an ensured bounded latency. For BE traffic, its latency and jitter performance degrade significantly (with increasing traffic) when the link overload is increased. The fully utilized link results in high collision possibility and unstable communication.

Our investigation in Fig. 7 is based on worst-case scenarios. In these experiments, we allocate 1.6Mbps, 18 Mb/s, and 30 Mb/s bandwidth to handle TTCAN, TT, and RC traffic, respectively. In Fig. 7, we evaluate the influence of the integration cycle on the performance of all traffic in a worst-case scenario. When reducing the period of the integration cycle, more overhead will be generated because of the PCF frames exchanged between the compression master and synchronization masters. On the other hand, a clock's drift is decreased with a shorter integration cycle and therefore improves the timing precision in the TT-E backbone network. It can be seen from Fig. 7(a) that TTCAN and TT traffic is not affected by the varying integration cycle because they are protected by link reservation. The throughput performance of RC and BE traffic is degraded due to the increased overhead. Fig. 7(b) demonstrates the jitter performance under the influence of the integration cycle. TTCAN, TT and RC traffic achieves a slight improvement in the jitter performance due to increased time precision when reducing the integration period. In contrast, BE traffic shows a worse jitter performance due to the dominant factor of increased overhead and packet loss rate.

## V. Conclusion

In this article, we present a time triggered communication network architecture that integrates two time-triggered protocols: namely TT-Ethernet and TT-CAN. Specifically, high speed TT-E is employed as the backbone network to comply with the high bandwidth requirement, while lower bandwidth TT-CAN is used as sub-networks for low speed ECUs with consideration of cost-effectiveness. Thanks to the synchronization and link reservation mechanism in both TT-E and TT-CAN, they are capable of supporting constant and low latency traffic with very low jitter. This guarantees reliability in-vehicle communication. Our investigation into fully overloaded links and worst-case scenarios further demonstrates the reliability and robustness of the integrated time-triggered communication network architecture.

### Acknowledgment

The authors would like to thank Dr. Ejaz Ahmed, a former NIST associate, for his involvement in the initial stages of this work.